%% file: main.tex
\documentclass[10pt,twocolumn,letterpaper]{article}
\usepackage{booktabs, colortbl} % 패키지 추가
\usepackage{subcaption}
\usepackage[pagenumbers]{iccv} % To force page numbers, e.g. for an arXiv version
\usepackage{svg}
\usepackage{algorithm}
\usepackage{algpseudocode}
\usepackage{multirow}

\usepackage{subcaption}  % subtable을 위한 패키지
\usepackage{booktabs} 


\definecolor{iccvblue}{rgb}{0.21,0.49,0.74}
\usepackage[pagebackref,breaklinks,colorlinks,allcolors=iccvblue]{hyperref}

\def\paperID{7774} % *** Enter the Paper ID here
\def\confName{ICCV}
\def\confYear{2025}

\title{Consistent Zero-shot 3D Texture Synthesis Using Geometry-aware Diffusion and Temporal Video Models}

\author{Donggoo Kang$^1$ \and Jangyeong Kim$^2$ \and Dasol Jeong$^1$ \and Junyoung Choi$^2$ \and Jeonga Wi$^2$ \and Hyunmin Lee$^1$ \and Joonho Gwon$^3$ \and Joonki Paik$^{1,*}$ \and 
$^1$Department of Image, Chung-Ang University \\
$^2$Graphics AI Lab, NCSOFT  \\
$^3$Department of Computer Science, University of Seoul \\
\tt\small \{dgkang, dasolj, hl, paikj\}@ipis.cau.ac.kr, \{jk, jc, jw\}@ncsoft.com}

\begin{document}
\twocolumn[{
\maketitle
\begin{center}
    \centering
    \captionsetup{type=figure}
    \includegraphics[width=\textwidth]{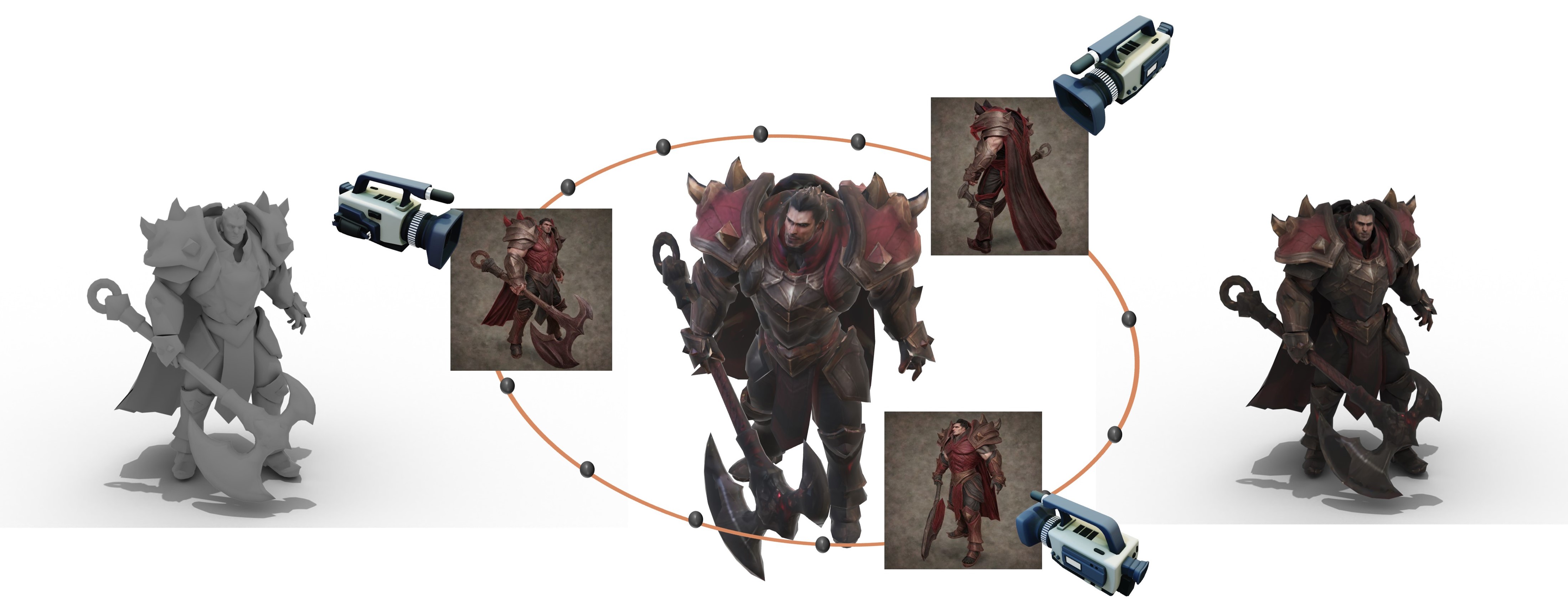}
    \captionof{figure}{This paper presents \textit{VideoTex}, a seamless texture synthesis approach that utilizes the temporal consistency of a video generation model. We also propose a component-based UV diffusion model to achieve semantic-aware UV diffusion.}
\end{center}
}]

\input{sec/0_abstract}    
\input{sec/1_intro}

\input{sec/2_relatedwork}
\input{sec/3_method}
\input{sec/4_experiment}
\input{sec/5_conclusion}

% WARNING: do not forget to delete the supplementary pages from your submission 

{
    \small
    \bibliographystyle{ieeenat_fullname}
    \bibliography{main}
}

\input{sec/X_suppl}

\end{document}

%% file: sec/0_abstract.tex
\begin{abstract} 
Current texture synthesis methods, which generate textures from fixed viewpoints, suffer from inconsistencies due to the lack of global context and geometric understanding. 
Meanwhile, recent advancements in video generation models have demonstrated remarkable success in achieving temporally consistent videos. 
In this paper, we introduce VideoTex, a novel framework for seamless texture synthesis that leverages video generation models to address both spatial and temporal inconsistencies in 3D textures. 
Our approach incorporates geometry-aware conditions, enabling precise utilization of 3D mesh structures. 
Additionally, we propose a structure-wise UV diffusion strategy, which enhances the generation of occluded areas by preserving semantic information, resulting in smoother and more coherent textures. 
VideoTex not only achieves smoother transitions across UV boundaries but also ensures high-quality, temporally stable textures across video frames. Extensive experiments demonstrate that VideoTex outperforms existing methods in texture fidelity, seam blending, and stability, paving the way for dynamic real-time applications that demand both visual quality and temporal coherence.
\end{abstract}

%% file: sec/1_intro.tex
\section{Introduction}

Generating high-quality textures for 3D models is a core challenge in computer graphics and computer vision, with direct implications for visual realism in applications like video games, virtual reality, and animated films. Textures are key to realistic perception, providing intricate surface details that enrich users' visual experience. However, achieving seamless textures across complex geometries while maintaining visual coherence from diverse viewpoints is particularly difficult. This challenge intensifies in dynamic models or environments where varying viewing angles can lead to visible artifacts, inconsistencies, and Janus problems.

\begin{figure*}[ht]
    \centering
    \includegraphics[width=\linewidth]{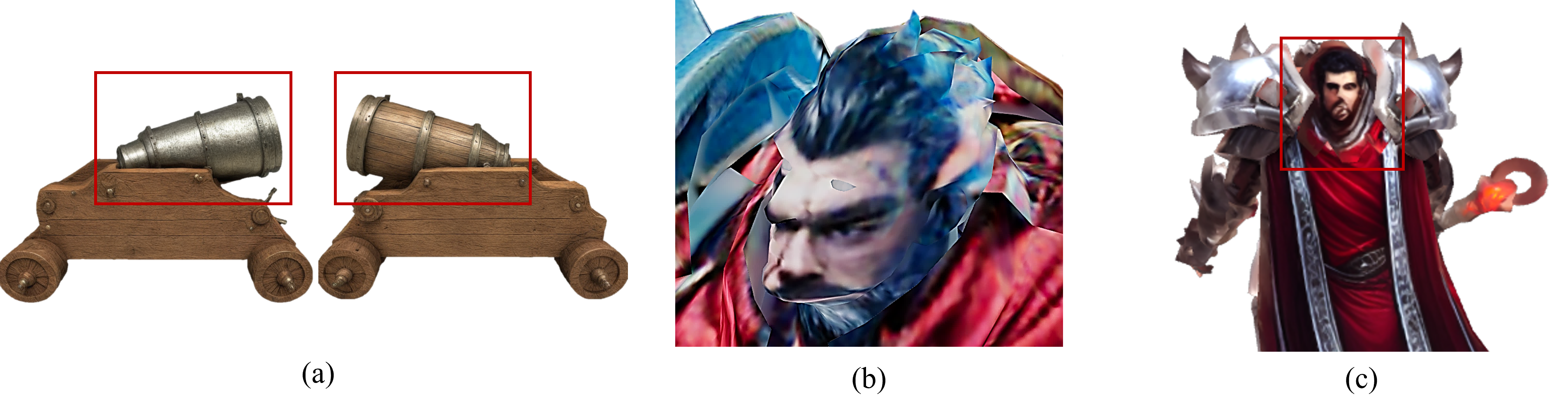}
    \caption{Limitations of existing texture synthesis methods: (a) lack of contextual awareness, (b) misalignment and lack of geometric consistency, and (c) the Janus problem.}
    \label{fig:problem}
\end{figure*}

Preceding texture synthesis methods~\cite{clipmesh, latent-nerf, text2tex, texfusion, texro, TEXTure, kim2024rocotex, MetaTextureGen, paint3d, flexitex} often suffer from limitations such as (a) lack of contextual awareness, (b) misalignment and lack of geometric consistency, and (c) the Janus problem. These issues stem from a reliance on fixed viewpoints, which constrains their ability to capture the \textbf{global 3D context} of models. As illustrated in Figure~\ref{fig:problem}
, these limitations frequently lead to spatial inconsistencies, including visible seams and distortions, especially at the boundaries where UV maps are stitched. Such artifacts become more pronounced when models are viewed from diverse angles or applied in dynamic, real-time applications. Additionally, occluded or hidden regions of the model often lack sufficient texture information, resulting in incomplete or distorted results that reduce the overall visual fidelity.

Recent advancements in video generation models\cite{wu2023tune, ho2022imagen, zhang2024show, blattmann2023align, wang2024magicvideo, girdhar2023emu, gao2024lumina, cong2023flatten, chen2024gentron, esser2023structure, ho2020denoising, neal2001annealed, jarzynski1997nonequilibrium}, which excel at capturing temporal dependencies between frames, provide a promising solution. These models have demonstrated the ability to ensure smooth transitions and consistent content across frames, making them highly effective for tasks that demand temporal stability. This inspires a novel approach of framing texture synthesis as a video generation problem, where each frame represents a different viewpoint or time step.

To address the dual challenges of spatial and temporal inconsistencies in texture synthesis, we propose a method that leverages video generation models for texture synthesis.
Our proposed method leverages the strengths of video generation models to address the temporal and spatial coherence challenges inherent in 3D texture synthesis. 
By conceptualizing texture synthesis as a video generation task, we map each component's texture generation process to a series of frames, where each frame represents a discrete viewpoint or temporal step across the surface of the 3D model.

Furthermore, a core challenge in 3D asset creation lies within the \textbf{UV mapping domain}, a critical process for converting 3D surfaces to 2D space for seamless texture application. However, UV mapping presents several challenges: (1) flattening intricate 3D shapes onto a 2D plane can introduce distortions and seams; (2) it causes a \textbf{loss of spatial coherence}, disrupting continuity in image processing; and (3) both artist-generated and automated UV maps struggle with these issues, making precise texture application difficult. Despite prior research, robust texture handling in the UV domain remains an open problem.

To address this, we observed that most 3D assets are built from reusable components. This insight forms the basis for a component-based approach in the UV domain, providing solutions for occlusion and ensuring seamless texture synthesis across complex geometries.

We propose a texture synthesis approach that operates in the UV domain by leveraging the component structure of 3D assets. Unlike traditional methods that generate a unified texture, our approach decomposes models into individual components, mapping each separately in UV space. This enables finer control over texture coherence and enhances the quality of synthesized textures.

Our main contributions are as follows:
\begin{itemize}
    \item We introduce \textbf{VideoTex}, a novel texture synthesis framework that formulates texture generation as a video generation problem, ensuring high temporal and spatial consistency.
    \item We employ \textbf{geometry-aware conditional generation} by leveraging 3D structure information, such as normal, depth, and edge maps, to align textures with complex geometries.
    \item We propose a \textbf{component-wise UV diffusion} strategy that enhances texture quality across UV seams and occluded regions, maintaining fine details and semantic coherence.
    \item Our method outperforms existing approaches in terms of texture fidelity, seam blending, and stability, paving the way for real-time applications that demand both high visual quality and dynamic adaptability.
\end{itemize}

%% file: sec/2_relatedwork.tex
\section{Related Work}

\subsection{3D Texture Synthesis}
Recent advancements in texture synthesis for 3D meshes integrate diffusion models with text-based guidance. Despite improvements, most methods rely on fixed viewpoint inference, limiting spatial coherence across perspectives—crucial for realistic 3D applications.

Text2Tex~\cite{text2tex} employs a depth-aware diffusion model to generate textures from multiple fixed viewpoints, mitigating view inconsistencies through dynamic segmentation. However, independent viewpoint treatment leads to visible seams and artifacts from untrained angles. Similarly, TexFusion~\cite{texfusion} uses a 3D-aware text-to-image model, but its fixed viewpoint approach causes texture misalignment on complex geometries.

RoCoTex~\cite{kim2024rocotex} enhances view consistency via symmetrical synthesis and regional prompts but struggles with seamless texture coherence across arbitrary viewpoints. Meta 3D TextureGen~\cite{MetaTextureGen} improves efficiency by conditioning a text-to-image model on 3D semantics in 2D space, yet its reliance on fixed viewpoints limits robustness in dynamic environments.

While these methods advance texture quality and consistency, their fixed viewpoint dependency restricts temporal stability. Our approach leverages video generation models to ensure seamless texture coherence across diverse perspectives.

\subsection{Video Generation Model}

Recent advancements in diffusion models have positioned them at the forefront of generative frameworks for both image and video synthesis, due to their scalability and enhanced training stability \cite{ho2020denoising, neal2001annealed, jarzynski1997nonequilibrium}. However, video generation entails unique complexities that surpass those of static image synthesis, primarily due to the inherent temporal dynamics and continuity required across frames. Existing methods often address this challenge by either adapting pre-trained image generation models through fine-tuning or by jointly training models to accommodate both images and videos \cite{wu2023tune, ho2022imagen, zhang2024show, blattmann2023align, wang2024magicvideo, girdhar2023emu, gao2024lumina, cong2023flatten, chen2024gentron, esser2023structure}. While effective, these approaches can limit video generation performance by imposing constraints inherited from the image-focused pre-training stage, potentially hindering the capture of smooth temporal transitions.

AnimateDiff, proposed by Guo et al.~\cite{guo2023animatediff}, advances text-to-image (T2I) diffusion models by integrating a motion module trained on real-world video data, enabling temporally consistent animations without fine-tuning the base T2I models. This modularity allows for diverse animated content, expanding T2I models from static images to dynamic video generation with minimal extra computation.

These developments emphasize the accelerated progress in video generation, particularly in embedding motion dynamics within diffusion frameworks. 

\subsection{Texture Representation in the UV Domain}

\begin{figure}[ht]
    \centering
    \includegraphics[width=\linewidth]{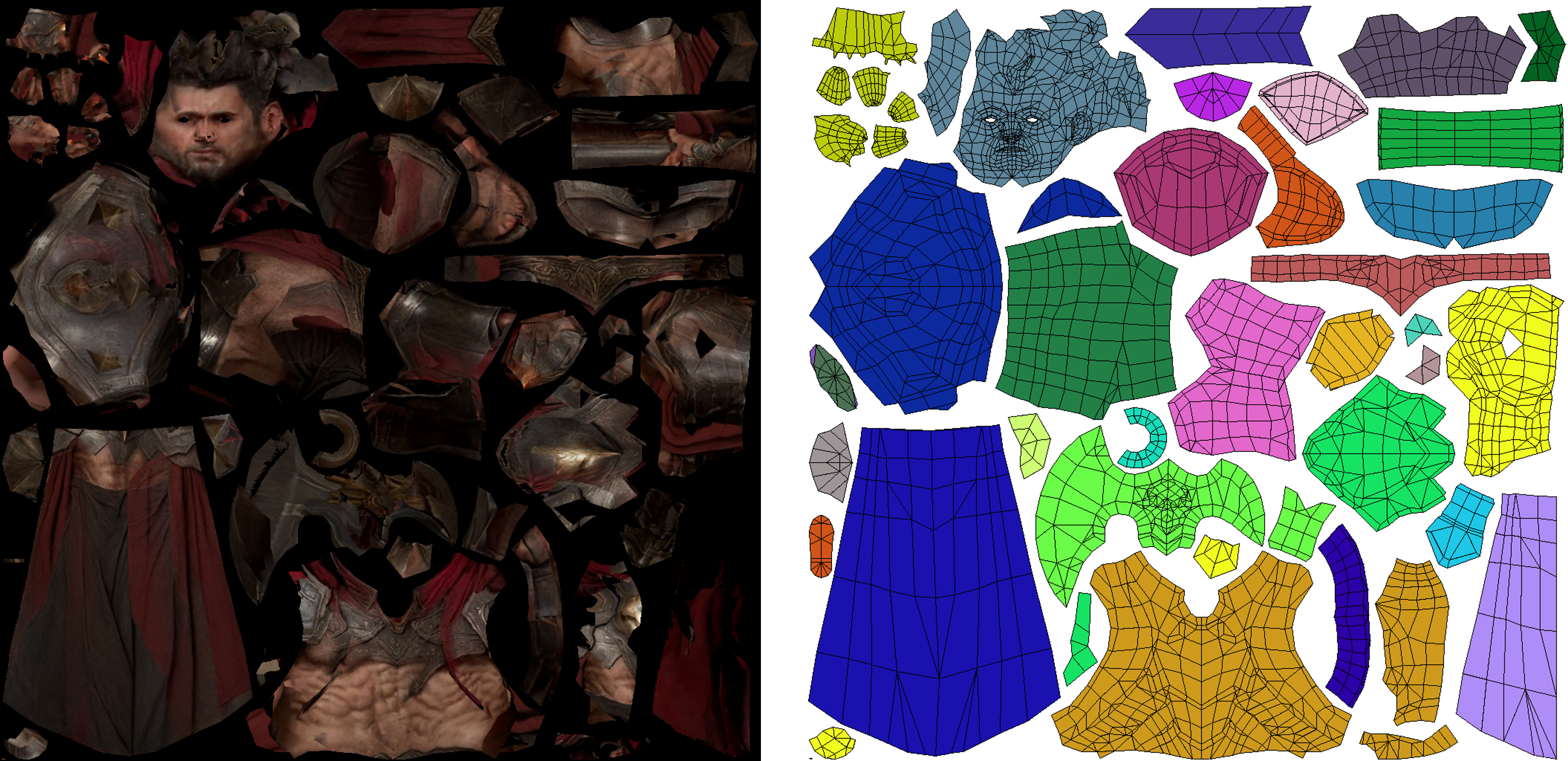}
    \caption{Visualization of the UV map for a 3D asset composed of multiple reusable components.}
    \label{fig:enter-label}
\end{figure}

\begin{figure}[h]
    \centering
    \begin{subfigure}{0.49\textwidth}
        \centering
        \includegraphics[width=\linewidth]{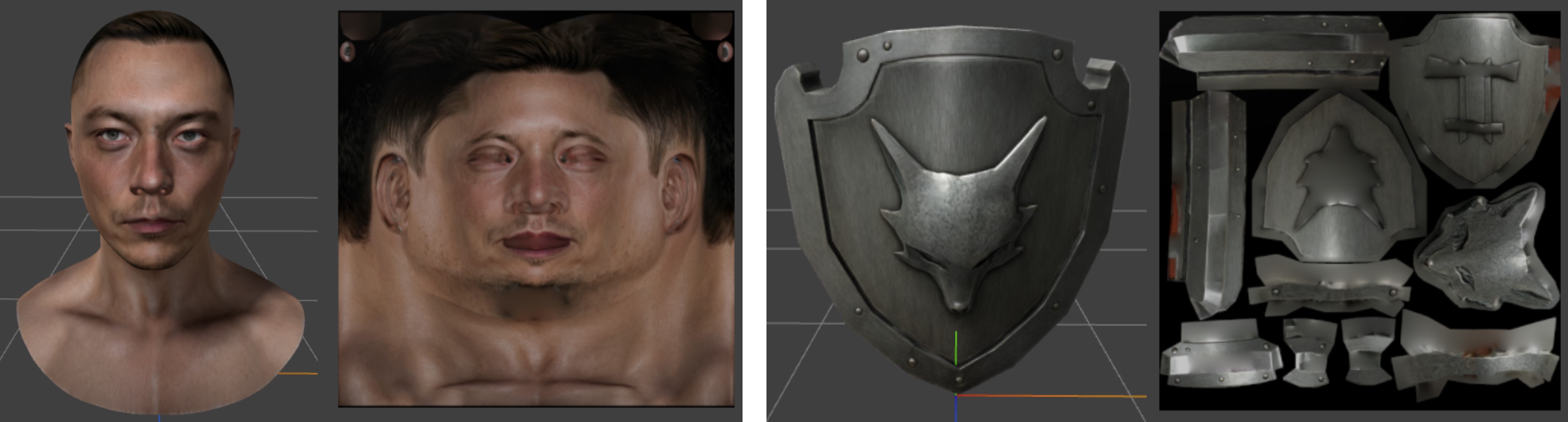}
        \caption{}
    \end{subfigure}
    \begin{subfigure}{0.49\textwidth}
        \centering
        \includegraphics[width=\linewidth]{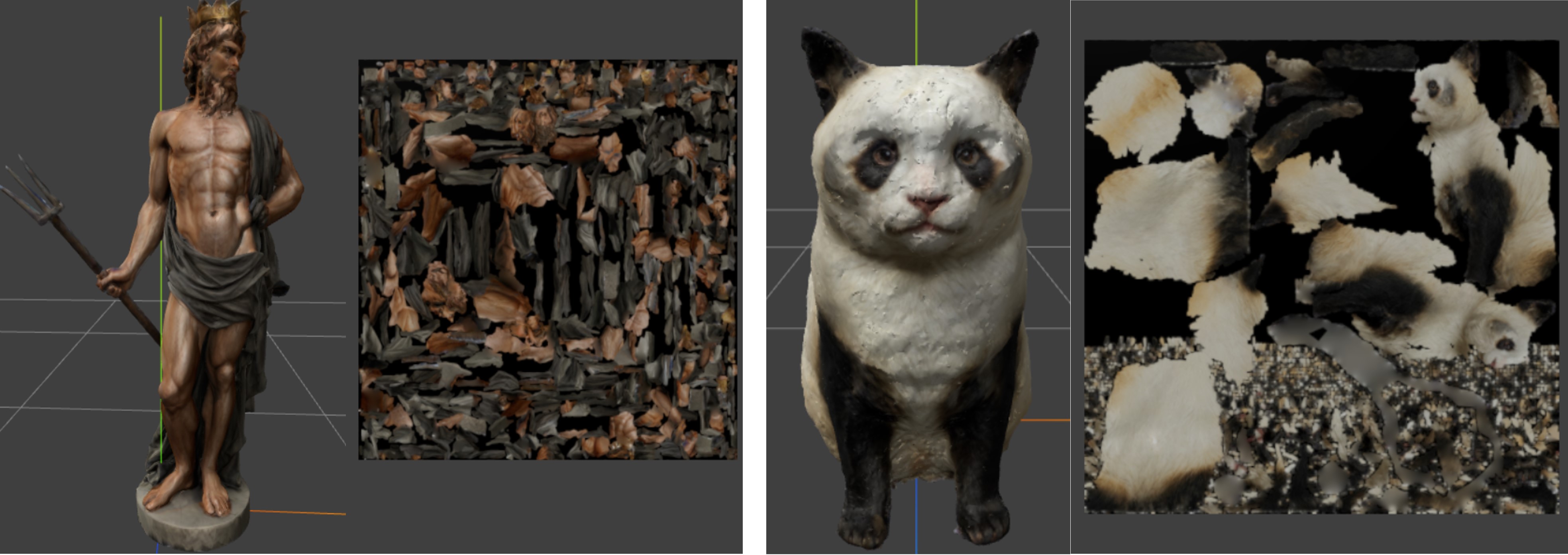}
        \caption{}
    \end{subfigure}
    \caption{UV map created by the 3D modeler (a) and UV map generated using an auto UV unwrap algorithm (b).}
    \label{fig:uv-process}
\end{figure}

\paragraph{UV Mapping}

UV mapping is a fundamental technique in 3D graphics for applying textures to mesh surfaces by mapping 2D coordinates \( (u, v) \) onto 3D vertices. Given a texture function \( T(u, v) \) and a mapping function \( M: \mathbb{R}^3 \rightarrow \mathbb{R}^2 \), the color at any point \( P \) on the 3D surface is:

\begin{equation}
    C(P) = T(M(P)),
    \label{eq:UVmapping}
\end{equation}

where \( M(P) = (u, v) \) represents the UV coordinates. This mapping enables detailed texturing and serves as a crucial method for preserving texture information in 3D assets. It also facilitates manual editing by artists, allowing them to refine textures directly in UV space.

However, as shown in Figure~\ref{fig:uv-process}, the structure of UV maps varies significantly depending on their creation method. In (a), manually authored UV maps by 3D modelers are often well-organized, maintaining spatial coherence that retains the locality of texture information. In contrast, (b) illustrates UV maps generated by automatic unwrapping algorithms, where spatial relationships in 3D space are disrupted due to irregular cuts and distortions. This loss of spatial information makes it challenging to leverage UV-domain features effectively for tasks like image-based processing.

Point-UV diffusion~\cite{point-uv} enhances geometric consistency in UV-space diffusion but struggles with fragmented UV cuts, leading to artifacts on complex shapes. Paint3D~\cite{paint3d} mitigates this using a position map, but its reliance on barycentric interpolation captures Euclidean rather than geodesic distances, causing semantic information loss and boundary distortions in 3D meshes.

\paragraph{Ill-Posed Nature of Reverse Projection}

Reverse projection maps 2D images generated by diffusion models onto 3D meshes, ensuring textures align with the model’s geometry. Mathematically, given a 2D texture image \( I \) and a 3D mesh \( M \) with vertices \( v_i \in \mathbb{R}^3 \), the projection function \( P^{-1} \) maps \( I \) onto \( M \):

\begin{equation}
    T(v_i) = I(P^{-1}(v_i)),
\end{equation}

where \( T(v_i) \) is the texture color at vertex \( v_i \), and \( P^{-1} \) maps each 3D vertex to its UV coordinate.

However, reverse projection is inherently an ill-posed problem, as multiple faces may map to the same pixel or a single face may receive multiple pixel values, leading to ambiguity and inconsistencies in texture reconstruction. This issue arises due to overlapping UV regions, resolution mismatches, and the lack of one-to-one correspondence between 2D texture space and the 3D surface. Addressing these challenges often requires post-processing techniques such as filtering, blending, or optimization-based refinement to ensure smooth and visually consistent texture mapping.

%% file: sec/3_method.tex
\begin{figure*}[ht]
    \centering
    \includegraphics[width=1\linewidth]{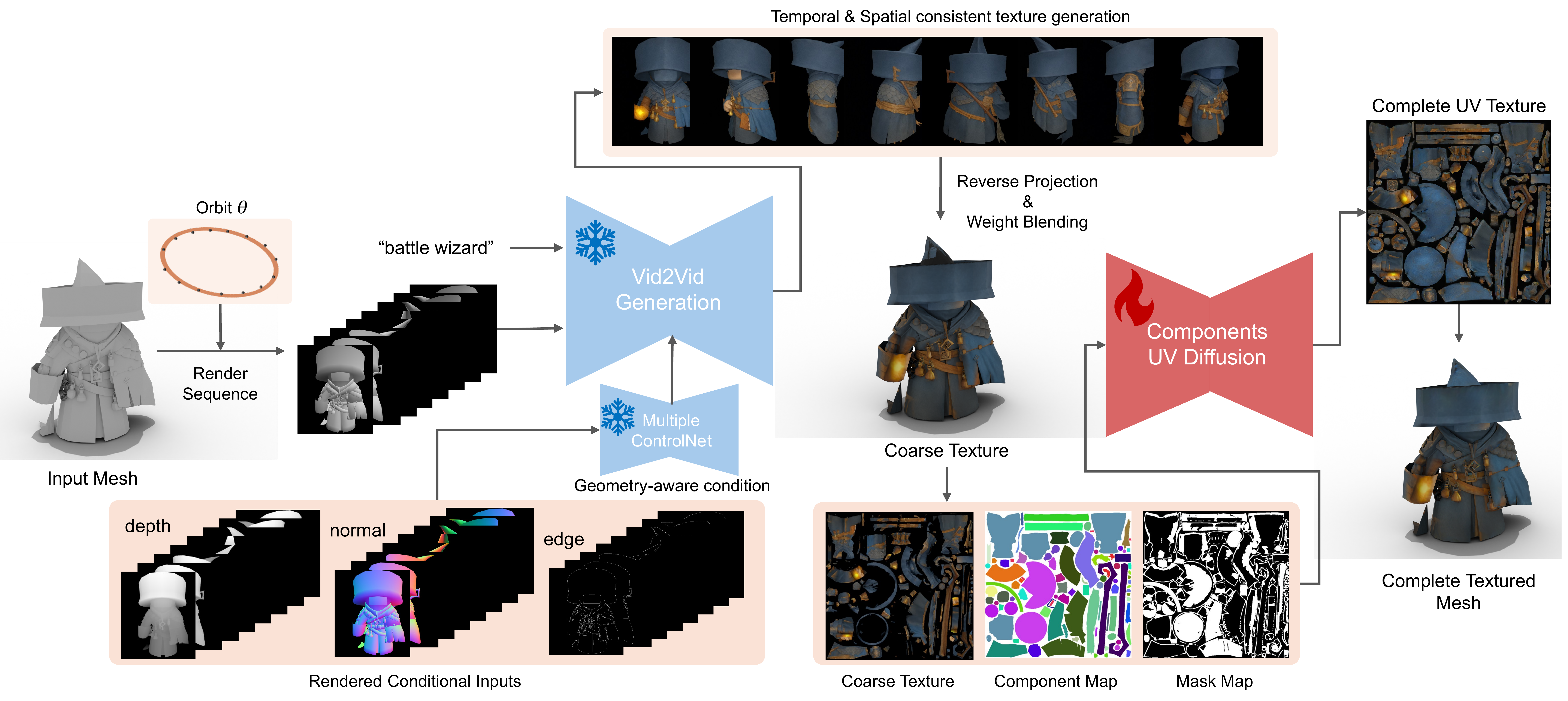}
    \caption{Overview of VideoTex. The proposed method generates temporally consistent textures in the coarse stage using a video generation model. In the refinement stage, occluded regions are refined through a structure-aware inpainting model.}
    \label{fig:architecture}
\end{figure*}

\section{Proposed Method}

\subsection{Overview of VideoTex Framework}

The overall pipeline of the VideoTex framework is illustrated in Figure~\ref{fig:architecture}. This framework ensures seamless texture synthesis with both spatial and temporal consistency through a two-stage process. First, video-based texture synthesis is performed using multiple ControlNets conditioned on normal, depth, and edge maps, generating geometrically aligned textures across viewpoints. Second, component-wise UV diffusion is applied to inpaint occluded regions in the UV domain, preserving semantic consistency. This approach enables high-quality, temporally stable textures across dynamic viewpoints.

\subsection{Geometry-Aware Conditional Generation}

To generate textures that align accurately with the structure of the 3D mesh, it is essential to condition the diffusion model on the geometry of the mesh. 
We leverage the SDXL~\cite{sdxl} model as our primary diffusion model, as it is well-suited for generating detailed and high-quality textures. 
To incorporate the 3D geometry effectively, we introduce three ControlNets~\cite{controlnet} conditioned on normal, depth, and edge maps derived from the rendered 3D mesh. These geometric cues enable the diffusion model to better understand and adhere to the underlying structure of the mesh, resulting in textures that are coherent and consistent with the 3D shape.

The conditioning process can be represented as follows:

\begin{equation}
T = G_\theta(z \mid G_{\text{normal}}, G_{\text{depth}}, G_{\text{edge}}),
\end{equation}
where \( T \) denotes the generated texture, \( G_\theta \) represents the generator network parameterized by \(\theta\), and \( z \) is a noise vector sampled from a prior distribution (e.g., Gaussian). \( G_{\text{normal}} \), \( G_{\text{depth}} \), and \( G_{\text{edge}} \) are the geometry features derived from the 3D mesh's normal, depth, and edge maps, respectively, serving as conditional inputs for the generator. 
By conditioning on these geometric features, the model learns to align textures with the 3D surface's intricate details, such as contours and depth variations, ensuring that the generated textures are faithful to the mesh's form.

\subsection{Temporal Consistency in Texture Generation with Video Models}

To ensure texture consistency across multiple viewpoints, we leverage the temporal coherence of video generation models. Conventional methods like text-to-video (T2V) and image-to-video (I2V) embed a single input at the start, limiting control over conditions throughout the video. Instead, we adopt a video-to-video (V2V) strategy, allowing dynamic conditioning at each stage.

Traditional fixed-viewpoint approaches require pre-processing for alignment, assuming predefined front, back, left, and right views. Our method eliminates this constraint by defining an orbit around the 3D mesh, generating a continuous 360-degree video sequence. This enhances temporal consistency by ensuring smooth transitions across viewpoints.

The orbit path is defined as:

\begin{equation}
O(t) = \left( r \cos \left( \frac{2 \pi t}{T} \right), r \sin \left( \frac{2 \pi t}{T} \right), z \right),
\end{equation}

where \( O(t) \) represents the 3D coordinates at time \( t \), \( r \) is the orbit radius, \( T \) is the video duration, and \( z \) is the camera height relative to the mesh. This approach captures a full 360-degree view with evenly spaced viewpoints.

At each viewpoint, we render the untextured mesh along with normal, depth, and edge maps, forming a sequential input video. Combined with a text prompt, this input is fed into the V2V model to generate temporally stable textures. The process is formulated as:

\begin{equation}
T_{seq}^\theta = G_\theta(V_{\text{mesh}} \mid C_{\text{normal}}, C_{\text{depth}}, C_{\text{edge}}, T_{seq}^{t-1,\theta}, p), 
\end{equation}

where \( T_{seq}^\theta \) is the output texture sequence, \( V_{\text{mesh}} \) is the input video, \( C_{\text{normal}}, C_{\text{depth}}, C_{\text{edge}} \) are condition maps, \( T_{seq}^{t-1,\theta} \) is the previous frame’s texture, and \( p \) is the text prompt.

By applying this V2V strategy with the orbit path, we achieve temporally consistent textures that transition smoothly across viewpoints, eliminating alignment concerns.

\subsection{Squared Confidence Texture Blending}

\begin{figure}[t]
    \centering
    \includegraphics[width=\linewidth]{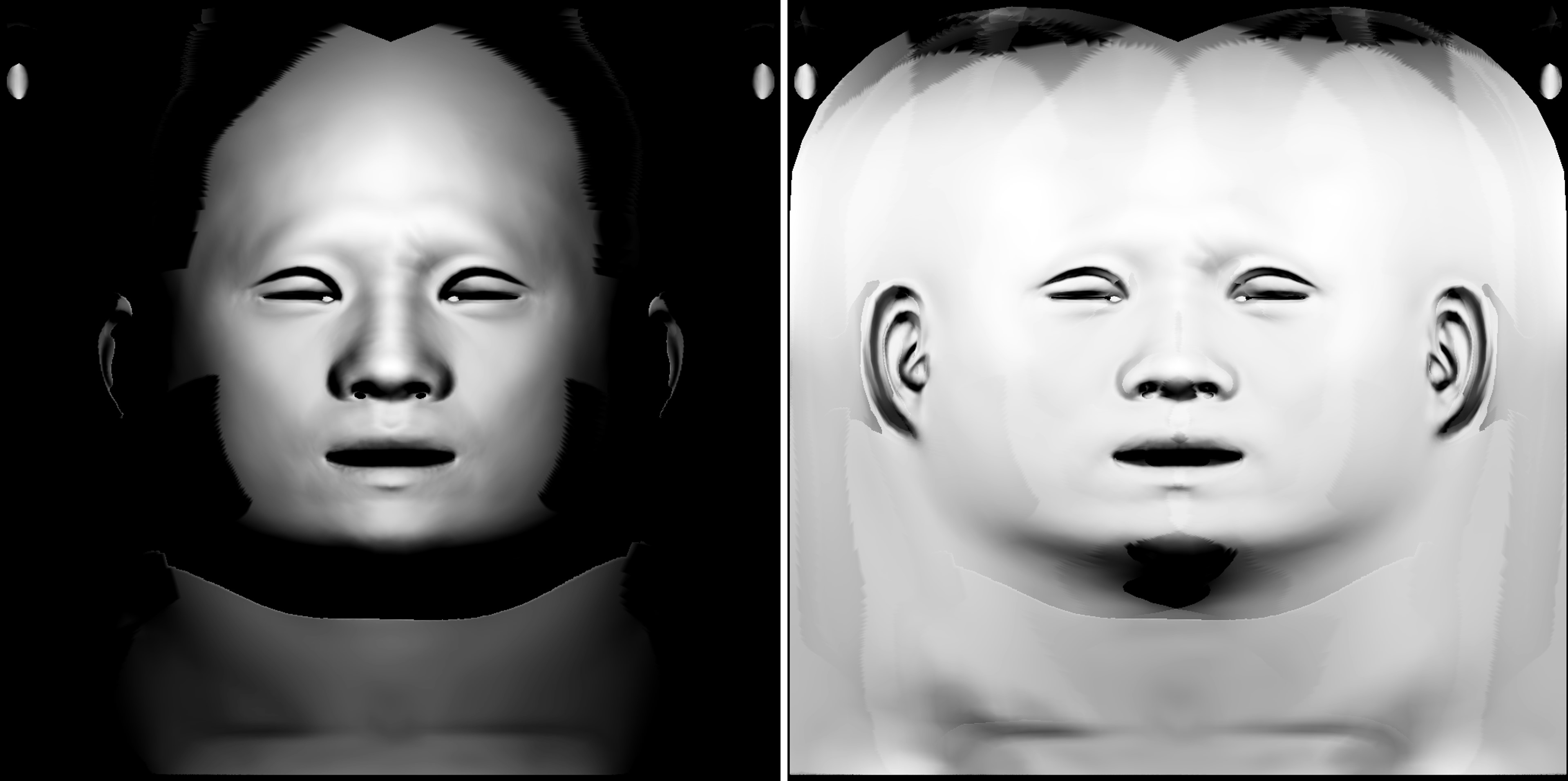}
    \caption{Confidence map from the first view (left) and the accumulated confidence map from 8 views (right).}
    \label{fig:confidence_1_8}
\end{figure}

To apply the generated video textures onto a 3D mesh, we perform reverse projection for each viewpoint, mapping each image frame onto the corresponding mesh areas to ensure texture alignment.

\begin{figure}
    \centering
    \includegraphics[width=\linewidth]{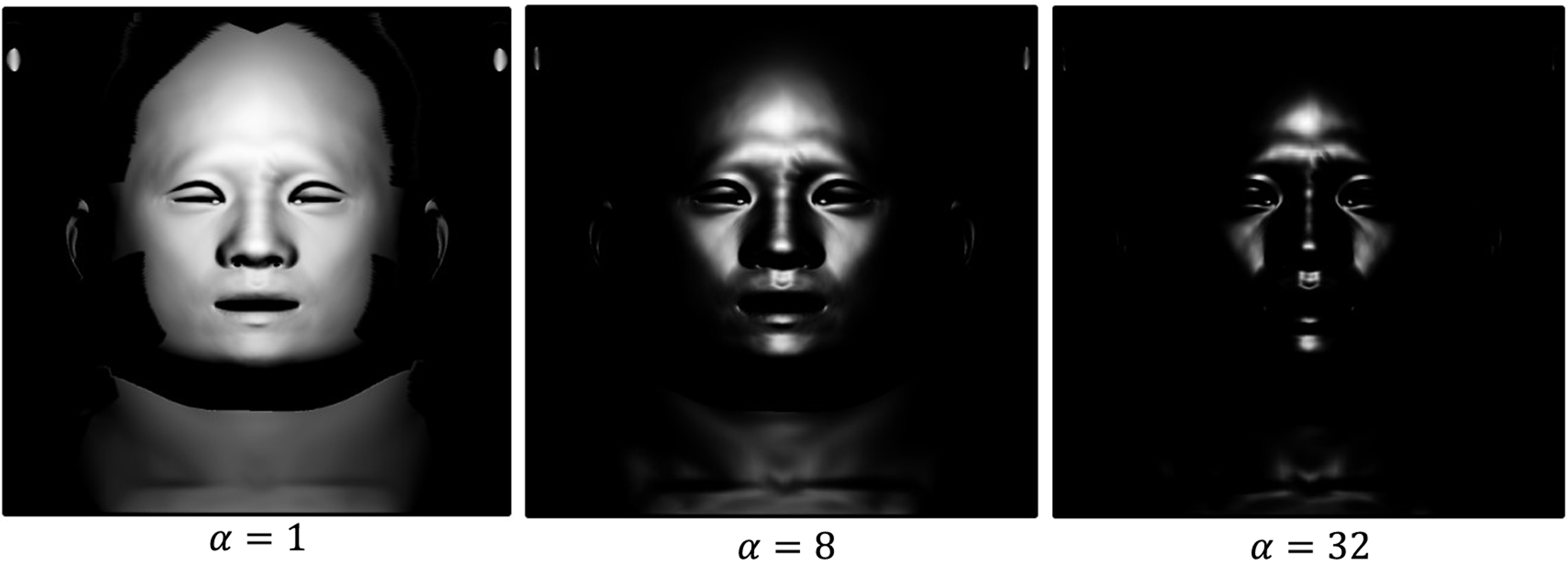}
    \caption{Visualization of confidence maps for square parameters.}
    \label{fig:confidence_1to32}
\end{figure}

Handling texels—pixels in texture space that map to 3D surface points—is crucial, as multiple texels from different viewpoints may project onto the same vertex, leading to over-smoothing (Figure~\ref{fig:confidence_1to32}, ~\ref{fig:confidence_vis}). To address this, we use a normal-based confidence weight:

\begin{equation}
w(v) = \cos(\theta) = \mathbf{n}_{v} \cdot \mathbf{n}_{\text{view}},
\end{equation}

where \( w(v) \) quantifies the alignment between a vertex normal \( \mathbf{n}_{v} \) and the viewing direction \( \mathbf{n}_{\text{view}} \), assigning higher confidence to viewpoints closely aligned with the surface.

Despite generating confidence maps (Figure~\ref{fig:confidence_1_8}), overlapping texel projections may cause inconsistencies. To mitigate this, we propose squared confidence texture blending:

\begin{equation}
w_{\text{final}}(v) = w(v)^\alpha.
\end{equation}

The final blended texture is computed as:

\begin{equation}
T_{\text{blend}}(v) = \frac{\sum_{i} w_{\text{final}}^{(i)}(v) \cdot T^{(i)}(v)}{\sum_{i} w_{\text{final}}^{(i)}(v)}.
\end{equation}

By adjusting the blending parameter \(\alpha\), we achieve a smoother, non-overlapping confidence map that enhances texture coherence across the mesh.

\begin{figure*}
    \centering
    \includegraphics[width=\linewidth]{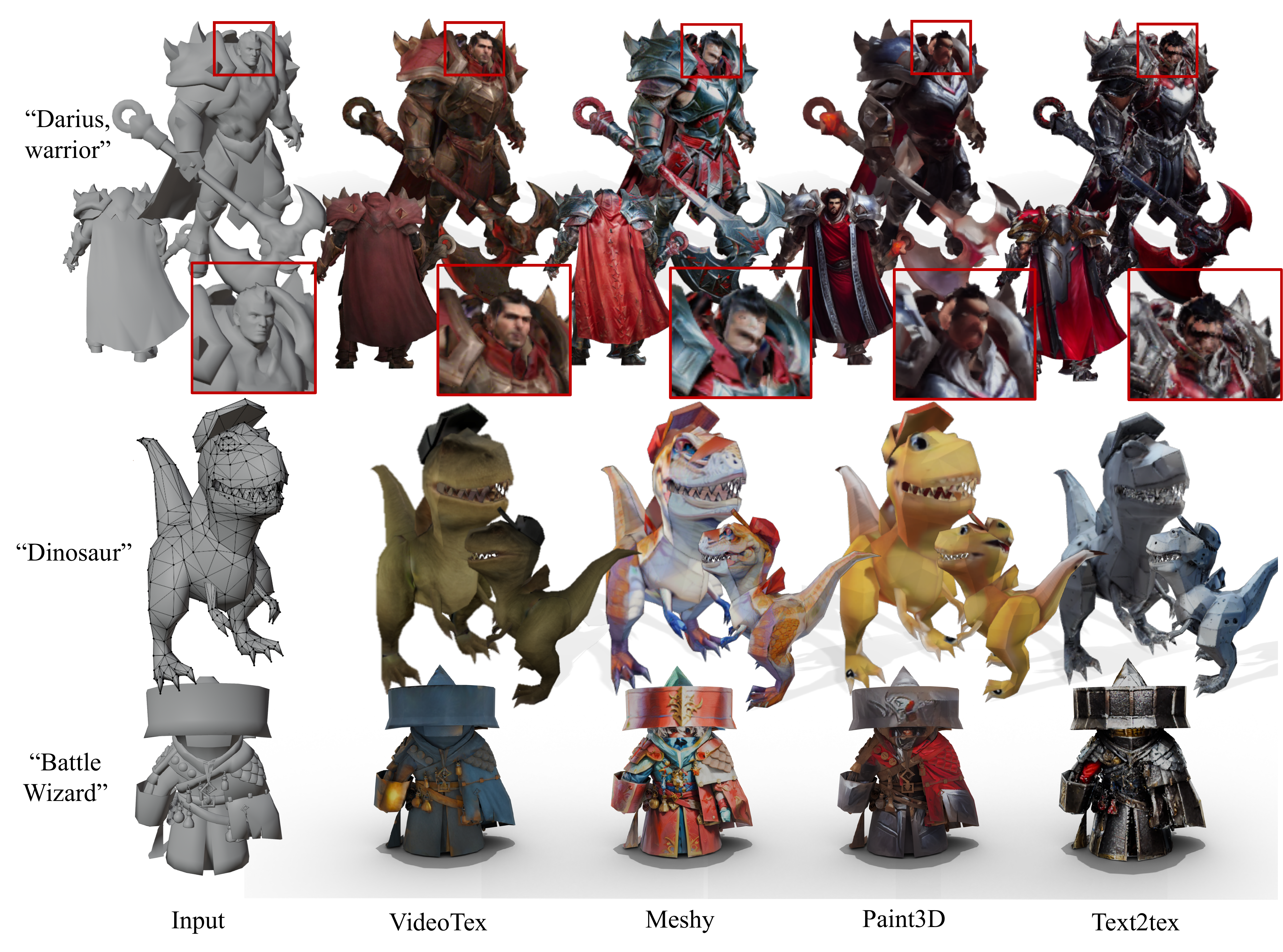}
    \caption{Qualitative comparison results with the state-of-the-art methods.}
    \label{fig:qualitative}
\end{figure*}

\subsection{Component-Wise UV Diffusion}

While video generation ensures temporal and spatial texture consistency, occluded regions remain a challenge. Although V2V methods capture broader views than fixed-viewpoint approaches, they struggle with deeply recessed or hidden areas. Inpainting in the UV domain can help, but as shown in Figure~\ref{fig:uv-process}, automated UV mapping often lacks spatial coherence, leading to discontinuities.

To address this, we propose \textbf{Component-UV Diffusion}, which divides the object into distinct components, allowing inpainting to be performed separately while preserving semantic coherence. This ensures that occluded regions receive consistent textures.

We trained a ControlNet tailored for component-wise UV inpainting using curated assets from Objaverse~\cite{objaverse}, segmented at the component level. The inpainting process is defined as:

\begin{equation}
T_{c} = D_\phi(T_{\text{input}} \mid C_{\text{component}}, M_{\text{mask}}),
\end{equation}

where \( T_{c} \) is the generated texture, \( D_\phi \) is the diffusion model, \( T_{\text{input}} \) is the partially rendered texture, \( C_{\text{component}} \) represents the component mask, and \( M_{\text{mask}} \) denotes untextured UV regions.

\begin{algorithm}
\caption{VideoTex}
\begin{algorithmic}[1]
\Require 3D Mesh $M$, UV Map $U$, Diffusion Model $G$, Video Generator $V$, ControlNets $C$
\Ensure Synthesized Texture $T$

\State \textbf{1. Render Geometry-Aware Condition Maps}
    \State \hspace{1em} $G_{\text{norm}}, G_{\text{depth}}, G_{\text{edge}} \gets \text{RenderConditionMaps}(M)$

\State \textbf{2. Video-Based Texture Synthesis}
    \State \hspace{1em} $V_{\text{input}} \gets \text{RenderSequence}(M, G_{\text{norm}}, G_{\text{depth}}, G_{\text{edge}})$
    \State \hspace{1em} $T_{\text{seq}} \gets V(V_{\text{input}}, C)$

\State \textbf{3. Reverse Projection and Blending}
    \State \hspace{1em} Project $T_{\text{seq}}$ onto $U$
    \State \hspace{1em} Compute confidence weights $w_{\text{final}}$
    \State \hspace{1em} Compute blend textures $T_{\text{blend}}$

\State \textbf{4. UV Inpainting for Occluded Regions}
    \State \hspace{1em} Identify occluded areas using mask $M_{\text{mask}}$
    \State \hspace{1em} Fill missing textures with UV diffusion: 
    \State \hspace{1em} $T_{\text{inpaint}} \gets G(U, M_{\text{mask}})$

\State \Return Final Texture $T$

\end{algorithmic}
\end{algorithm}

%% file: sec/4_experiment.tex
\section{Experiments}

\begin{table*}[ht]
    \centering
    \caption{Quantitative comparison results}
    \label{tab:results}
    \begin{subtable}{0.59\textwidth}  % 첫 번째 표
        \centering
        \caption{Comparison of different methods in texture synthesis}
        \begin{tabular}{cccccc}
        \toprule
        \multirow{2}{*}{Method} & \multirow{2}{*}{KID $\downarrow$} & \multicolumn{3}{c}{User Study (\%)} & \multirow{2}{*}{Time $\downarrow$}\\
        \cmidrule(lr){3-5}
          & & Quality $\uparrow$ & Consistency $\uparrow$ & Alignment $\uparrow$ \\
        \midrule
        Text2Tex & 9.44 & 4.0 & 0.0 & 4.0 & 446  \\
        \cmidrule(lr){1-1} \cmidrule(lr){2-2} \cmidrule(lr){3-5} \cmidrule(lr){6-6}
        Paint3D & 6.83 & 16.0 & 6.0 & 20.0 & 36 \\
        \cmidrule(lr){1-1} \cmidrule(lr){2-2} \cmidrule(lr){3-5} \cmidrule(lr){6-6}
        Meshy & 6.13 & 26.0 & 4.0 & 18.0 & 50\\
        \midrule
        \textbf{VideoTex} & \textbf{4.61} & \textbf{54.0} & \textbf{90.0} & \textbf{58.0} & \textbf{32} \\
        \bottomrule
        \end{tabular}
    \end{subtable}
    \hfill
    \begin{subtable}{0.39\textwidth}  % 두 번째 표
        \centering
        \caption{Effect of frame rate on performance}
            \begin{tabular}{ccc}
            \toprule
            Frame rate & KID $\downarrow$ & Time \\
            \midrule
            4 & 4.85 & 19 \\
            \cmidrule(lr){1-1} \cmidrule(lr){2-2} \cmidrule(lr){3-3}
            8 & \textbf{4.61} & 32 \\
            \cmidrule(lr){1-1} \cmidrule(lr){2-2} \cmidrule(lr){3-3}
            16 & 4.70 & 53  \\
            \cmidrule(lr){1-1} \cmidrule(lr){2-2} \cmidrule(lr){3-3}
            24 & 4.72 & 72 \\
            \bottomrule
            \end{tabular}
    \end{subtable}
\end{table*}

\subsection{Implementation Details}

We utilized the StableDiffusionXL (SDXL)\cite{sdxl} model along with normal, depth, and canny ControlNets\cite{controlnet} as our generation backbone. For rendering normal and depth maps from the mesh, we employed the Trimesh library, while edge maps were generated using Canny edge detection. Texture generation was managed through the ComfyUI framework, and UV diffusion was implemented using the Diffusers library. All network inputs were resized to 1024 $\times$ 1024. For UV diffusion model training, a single A100 GPU with 40GB memory was used, with inference conducted on an RTX 4090 GPU. We used a frame rate of 8 to generate the video, with $\alpha$ set to 8. The control strengths were set to 0.7 for the depth, normal, and edge ControlNets, and 0.5 for UV Component ControlNet.

\begin{figure*}[ht]
    \centering
    \includegraphics[width=\linewidth]{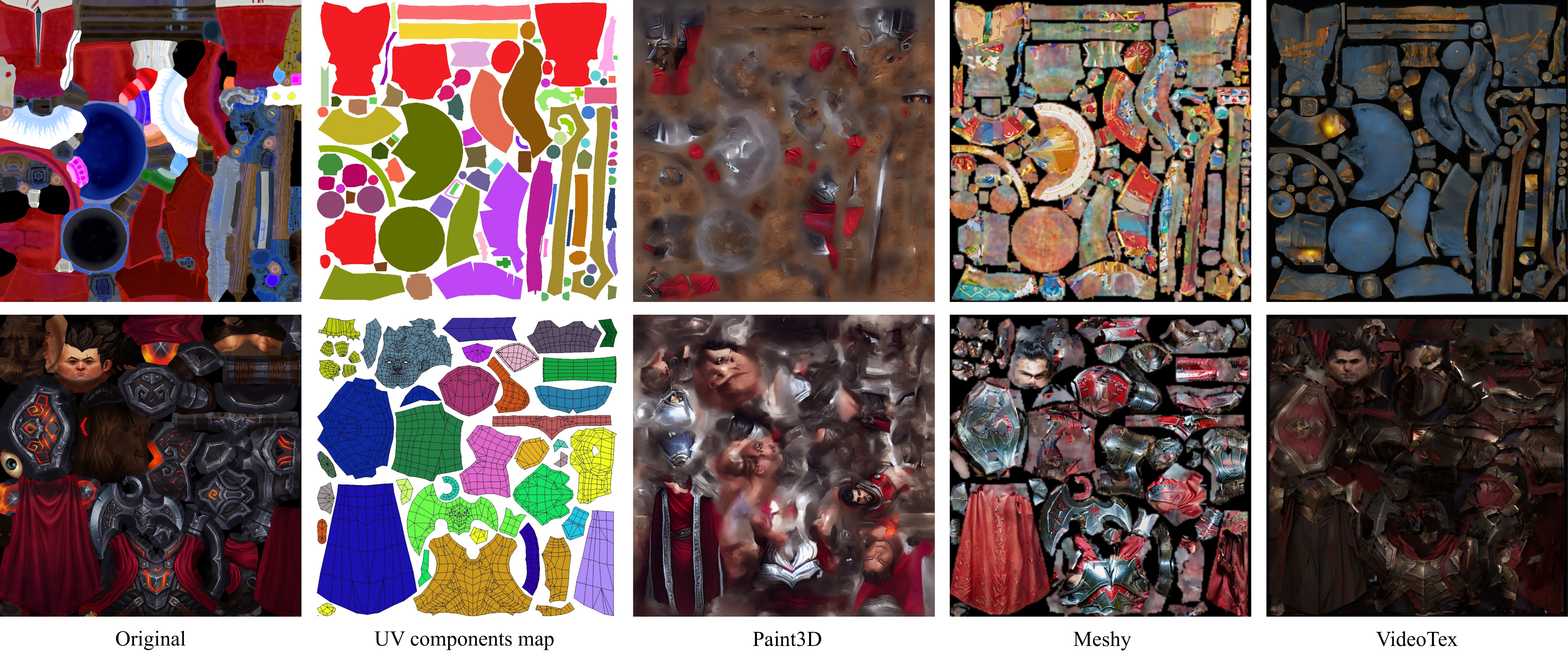}
    \caption{UV texture map comparison results}
    \label{fig:UVtexture_comparison}
\end{figure*}

\begin{figure}
    \centering
    \includegraphics[width=\linewidth]{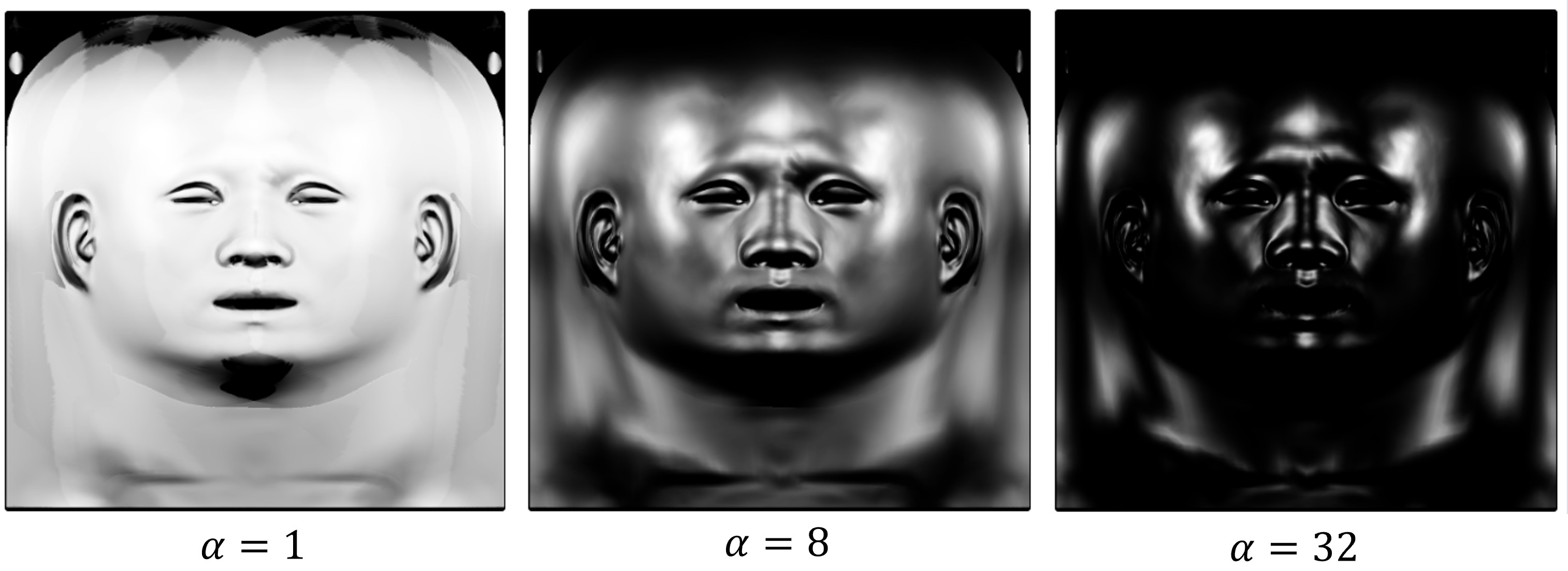}
    \caption{Visualization of accumulated confidence maps from 8 views with varying square parameters.}
    \label{fig:confidence_1to32}
\end{figure}

\begin{figure}
    \centering
    \includegraphics[width=\linewidth]{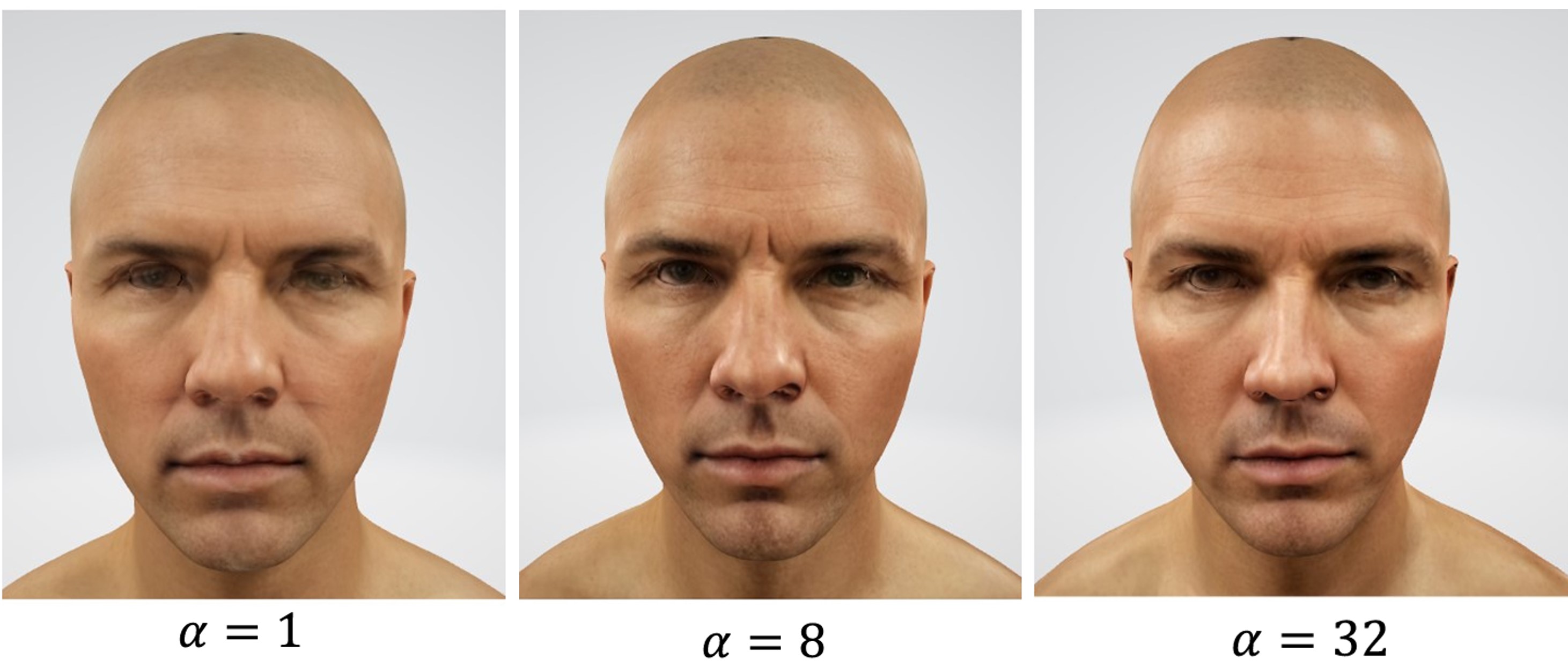}
    \caption{Texture results by square parameter \(\alpha\).}
    \label{fig:confidence_vis}
\end{figure}

\paragraph{Dataset}

For training the UV Diffusion model, we selected a subset from the Objaverse~\cite{objaverse} dataset. Through a filtering process that excluded overly simplistic and highly intricate models, we curated a collection of 37,979 meshes. Text prompts were sourced using the Cap3D~\cite{cap3d} method for prompt generation.

\subsection{Results and Comparisons}

We experiment with SeamlessTex using various texturing methods. 
However, due to limited publicly available source code, we compare VideoTex against only Text2tex~\cite{text2tex}, Paint3D~\cite{paint3d}, and Meshy.

\paragraph{Qualitative Comparison}

Figure~\ref{fig:qualitative} compares the original texture of the input with the results from Paint3D and Text2Tex, and Meshy. 
For the ``Darius" asset, we observe that the face and back cape regions are rendered naturally. 
In the case of the ``Dinosaur" asset, it successfully produces a consistent texture. 
Likewise, for the ``Battle Wizard" asset, our method generates a geometry-aware and consistent texture.
For a more detailed comparison, please refer to the supplementary material.

\paragraph{Quantitative Comparison}

The generated textures are evaluated using Kernel Inception Distance (KID), which is a commonly used image quality and diversity metric for generative models. 
Table~\ref{tab:results} shows that VideoTex achieves the lowest KID score, indicating higher quality and diversity of the generated images. 

Since there are no established metrics for evaluating 3D quality, we conducted a human evaluation similar to methods used in other studies. We randomly presented textured meshes to 30 participants and assessed them across three criteria: quality, consistency, and alignment. As shown in Table~\ref{tab:results}, the results demonstrated that our proposed method received high scores across these categories.

\subsection{Ablation Study}

\paragraph{Framerate and Texture Consistency}

Our vid2vid model, trained to generate 24-frame videos, faces increased generation time as frame count rises, so we tested 4, 8, 16, and 24 frames. As seen in Table~\ref{tab:results}, KID scores are similar for 8 frames and above, showing that higher framerates don’t always improve quality. This is due to reduced viewpoint variation at higher framerates, which, even with our blending method, leads to pixel overlap and an over-smoothing effect on texels.

\paragraph{Squared Confidence Blending Effectiveness}

Figure~\ref{fig:confidence_vis} shows the experimental results for parameter \(\alpha\) which controls the strictness of applying the confidence map for each view. Similar to Figure~\ref{fig:confidence_1to32}, only distinct textures from each view are applied, resulting in sharper outputs as the value of \(\alpha\) increases.

\subsection{Discussion and Limitations}
Recent video generation models like Sora~\cite{OpenAI2024Sora}, Kling, and Veo2~\cite{DeepMind2025Veo2} achieve superior temporal consistency but are closed-source and computationally expensive, making them impractical for our framework. Instead, we use AnimateDiff, the only open-source model compatible with our SDXL-based pipeline, balancing efficiency and quality.

While VideoTex improves temporal and spatial consistency, some limitations remain. AnimateDiff's quality lags behind state-of-the-art video models, and our component-wise UV diffusion may struggle with highly fragmented UV mappings. Future work will refine UV diffusion techniques and explore more advanced video models to enhance texture quality and coherence.

%% file: sec/5_conclusion.tex
\section{Conclusion}

This paper introduces \textit{VideoTex}, a pioneering approach in 3D texture synthesis that addresses long-standing challenges in the field, including spatial and temporal coherence, UV seam handling, and occlusion management. By reframing texture synthesis as a video generation problem, VideoTex leverages video diffusion models to produce seamless textures that maintain visual consistency across dynamic viewing angles. Our geometry-aware conditional generation, combined with a novel component-based UV diffusion strategy, enhances fidelity at the boundaries and provides realistic texture for occluded areas, outperforming existing methods both quantitatively and qualitatively.
Experimental results demonstrate VideoTex’s robustness in achieving fine texture alignment, high temporal stability, and seamless transitions. 
The combination of component-wise inpainting with squared confidence texture blending uniquely positions VideoTex to produce cohesive, artifact-free textures across complex geometries.

% We introduced VideoTex, a novel 3D texture synthesis approach that utilizes video generation models to achieve seamless spatial and temporal consistency. By integrating geometry-aware conditional generation with a component-wise UV diffusion strategy, our method effectively addresses challenges related to UV seam transitions and viewpoint stability, outperforming existing techniques in maintaining high-quality textures across complex geometries, including occluded areas. Experimental results demonstrate VideoTex’s superior performance in texture fidelity, seam smoothness, and temporal stability, making it ideal for real-time applications in gaming, virtual reality, and animation. This work paves the way for further advancements in dynamic and interactive environments, where both visual quality and processing efficiency are essential.

%% file: sec/X_suppl.tex
\maketitlesupplementary
% \onecolumn
\renewcommand{\thesection}{\Alph{section}}

\section{UV Dataset Details}

To develop the component UV diffusion model, we curated a carefully designed subset of the Objaverse dataset. The asset selection process involved filtering based on the number of faces and vertices to eliminate objects that were either excessively simplistic or overly complex, ensuring a balanced dataset that represents a wide range of geometric structures. Additionally, scanned datasets were excluded due to the prevalence of automated UV unwrapping algorithms in their UV map generation, which often lack the intentional semantic structure needed for this study. Instead, the dataset was meticulously curated to include assets that were manually crafted by professional 3D modelers, reflecting higher levels of design intention and semantic relevance.

% \begin{figure}
%     \centering
%     \includegraphics[width=0.4\linewidth]{figures/pieChart.png}
%     \caption{Distribution of components. Most assets are composed of multiple components.}
%     \label{fig:enter-label}
% \end{figure}

\begin{table}[h]
\setlength{\tabcolsep}{14pt}
\centering
\begin{tabular}{@{}lrr@{}}
\toprule
\textbf{Category}       & \textbf{Count} & \textbf{Proportion} \\ \midrule
1    & 12,419  & 32.7\%    \\
1-10 & 18,534  & 48.8\%    \\
10+  & 7,026  & 18.5\%    \\ \midrule
\textbf{Total}          & 37,979               & 100\%            \\ \bottomrule
\end{tabular}
\caption{Distribution of frequencies for components of meshes.}
\label{tab:freq}
\end{table}

The resulting dataset comprises 37,979 assets, with the detailed distribution of components provided in Table~\ref{tab:freq}. As shown in the table, a substantial proportion of the assets are composed of one or more reused components. These components are critical as they represent the sole carriers of semantic information within the UV maps. This unique characteristic is further illustrated in Figure~\ref{fig:UVs}, where the UV texture map is displayed alongside the corresponding components map. The figure underscores the ability of these components to encode distinct and meaningful \textbf{semantic} information within the UV domain, reinforcing their importance in the proposed model's training and evaluation process.

\section{UV map Comparison Results}
We conducted a comprehensive comparative experiment to evaluate the effectiveness of UV component diffusion by visualizing its impact within the UV domain. This experiment was designed to highlight the semantic-awareness of our approach in addressing the inpainting task. As depicted in Figure~\ref{fig:UVtexture_comparison}, our method demonstrates a clear advantage over existing approaches by producing inpainting results that not only preserve the semantic integrity of the UV domain but also exhibit superior alignment with the underlying geometric structure. These findings underscore the robustness and effectiveness of our approach in generating contextually coherent textures within the UV space.

\section{Further Experimental Results}

To comprehensively evaluate the performance of the proposed VideoTex framework, we conducted additional qualitative experiments to demonstrate its capability in texture generation. Figure 3 showcases the results of applying different prompts to the same input mesh, illustrating the versatility of our method. As observed in the figure, VideoTex effectively captures and reflects the intricate geometry of the mesh while generating textures that align closely with the semantic context of each prompt. This highlights the method's ability to integrate geometric fidelity with prompt-driven texture synthesis, further validating its effectiveness in handling diverse texture generation scenarios.

\section{User Study Details}

Metrics such as Fréchet Inception Distance (FID) are widely utilized for evaluating the quality of 2D images; however, their applicability to 3D contexts is inherently limited. In the case of 3D assets, evaluations are typically performed by rendering the objects as 2D images from specific viewpoints, such as front, back, and side perspectives, and subsequently comparing these renderings. While this approach provides some insights, it fails to comprehensively assess critical attributes such as texture consistency and diversity across the entirety of a 3D object. These limitations underscore the need for a more robust evaluation framework tailored to the unique challenges of 3D asset assessment.

To address these challenges, we designed and conducted a user study to evaluate the proposed methods more holistically. The user study was meticulously structured to include questions focusing on three key aspects of 3D asset quality: \textbf{texture quality, consistency, and geometric fidelity}. To facilitate an interactive and immersive evaluation experience, we developed dynamic HTML pages incorporating 3D viewers, enabling participants to manipulate the assets interactively in real time. This setup allowed users to rotate, zoom, and closely inspect the assets from various angles, ensuring a thorough examination of their characteristics.

The study involved 30 professional 3D modelers, who participated as evaluators, lending their domain expertise to provide informed assessments. To ensure the fairness and objectivity of the evaluation process, we anonymized the names of the methods under comparison and randomized the presentation order of the assets. This rigorous evaluation setup was applied to a total of 20 assets, yielding valuable insights into the relative performance of the methods in a real-world, user-driven context.

\begin{figure*}
    \centering
    \includegraphics[width=\linewidth]{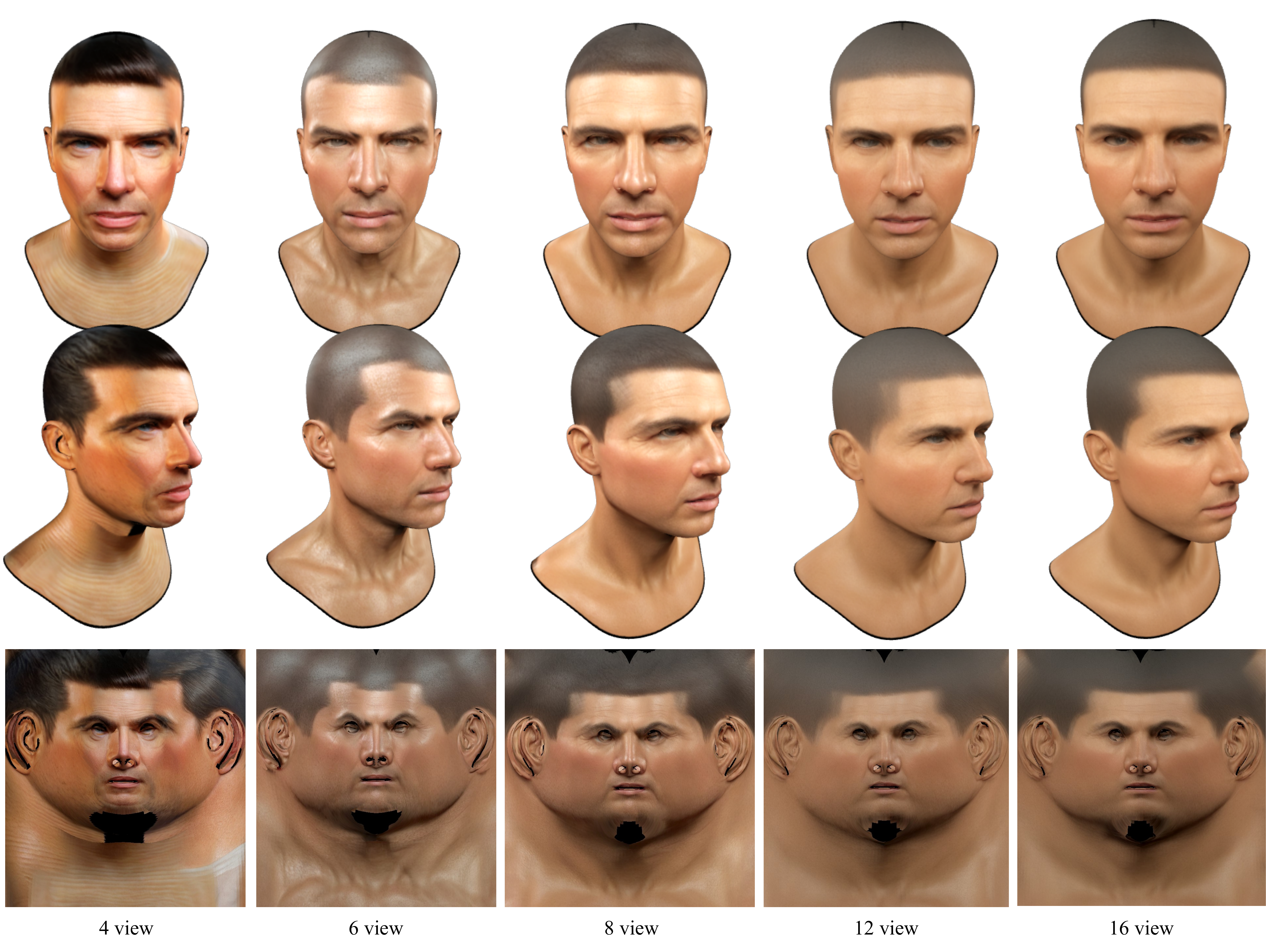}
    \caption{UV texture map comparison results}
    \label{fig:tom_views}
\end{figure*}

\begin{figure*}
    \centering
    \includegraphics[width=\linewidth]{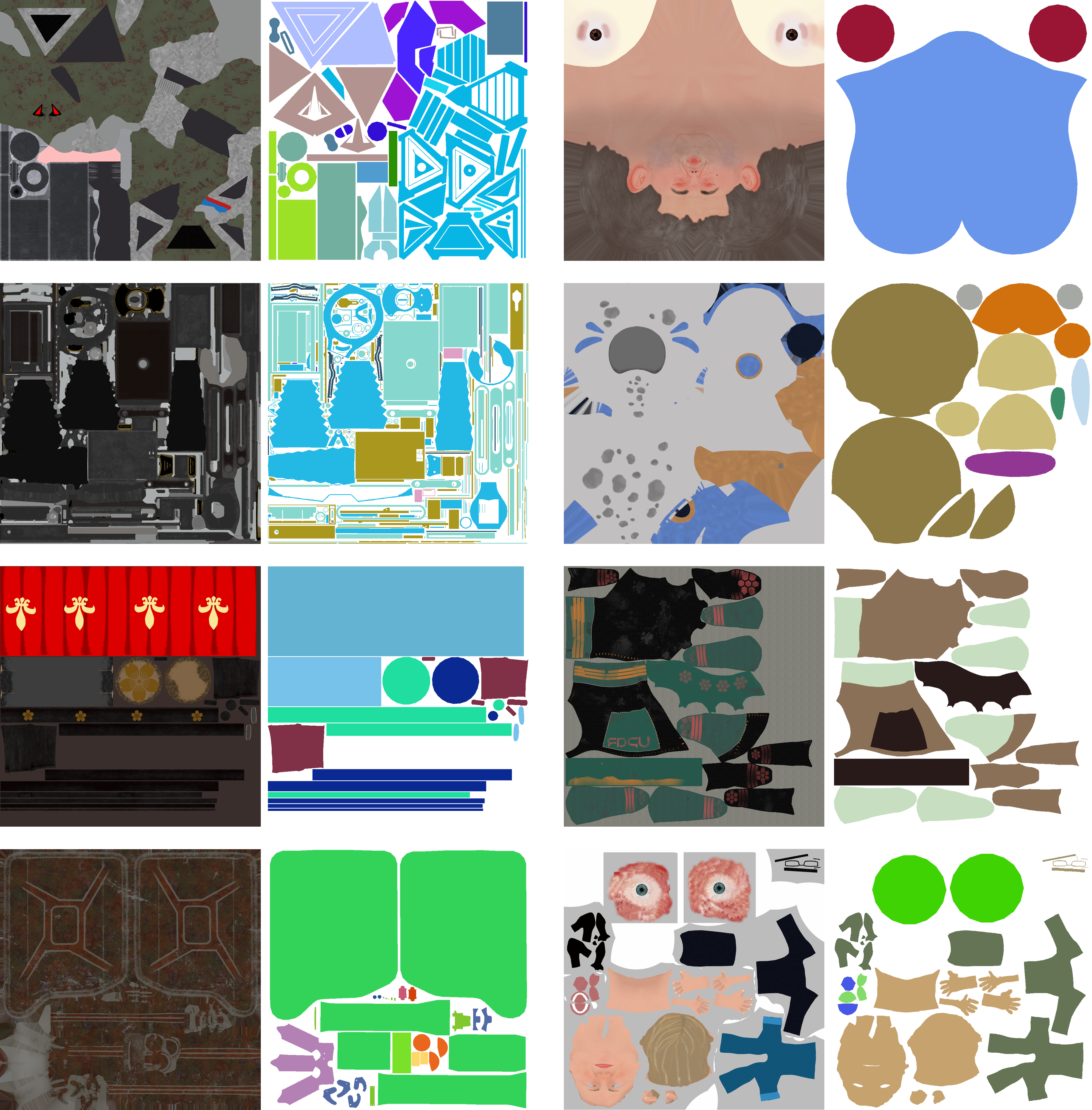}
    \caption{UV texture map and its corresponding UV component map from the training dataset}
    \label{fig:UVs}
\end{figure*}

\begin{figure*}
    \centering
    \includegraphics[width=\linewidth]{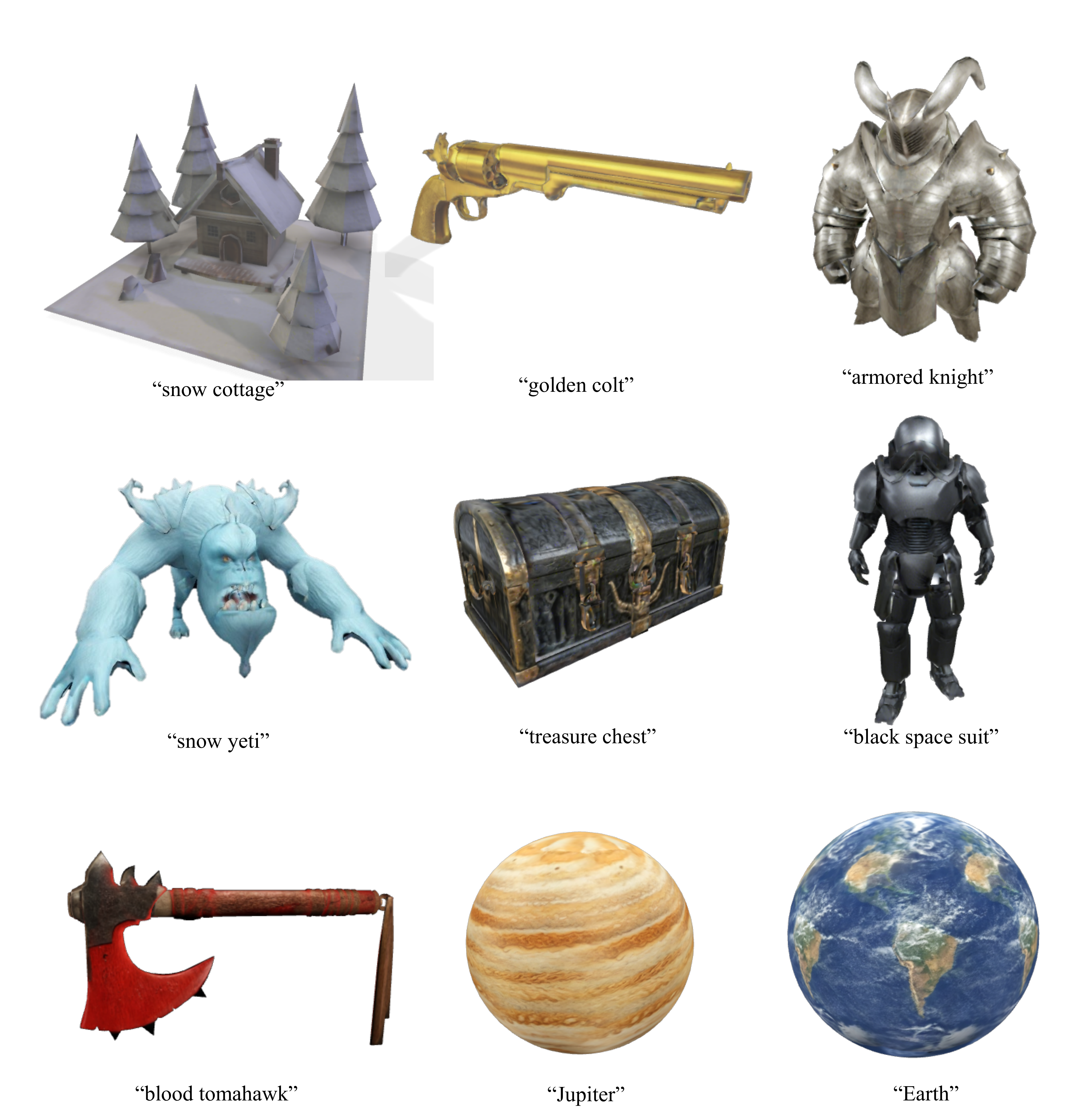}
    \caption{Additional experimental results}
    \label{fig:enter-label}
\end{figure*}